# Optimizing the half-gcd algorithm[*]

Joris van der Hoeven

CNRS, École polytechnique, Institut Polytechnique de Paris
Laboratoire d'informatique de l'École polytechnique (LIX, UMR 7161)
1, rue Honoré d'Estienne d'Orves
Bâtiment Alan Turing, CS35003
91120 Palaiseau, France

*Email:* vdhoeven@lix.polytechnique.fr

*December 23, 2022*

In this paper, we propose a carefully optimized "half-gcd" algorithm for polynomials. We achieve a constant speed-up with respect to previous work for the asymptotic time complexity. We also discuss special optimizations that are possible when polynomial multiplication is done using radix two FFTs.

## 1. Introduction

The computation of greatest common divisors is the key operation to be optimized when implementing a package for rational number arithmetic. For integers of small bit-length $n$, one may use Euclid's algorithm, which has a quadratic time complexity $O(n^2)$. Asymptotically faster algorithms were first proposed by Knuth [20] and Schönhage [29], based on earlier ideas by Lehmer [22]. Schönhage's algorithm has a logarithmic $\asymp \log n$ overhead with respect to integer multiplication, which is believed to be asymptotically optimal. Many variants and improvements have been proposed since, mainly aiming at faster practical implementations [36, 32, 23, 26, 4, 27]. All these subquadratic algorithms are based on a recursive reduction to half of the precision; for this reason it is convenient to regroup them under the name "half-gcd algorithms".

An analogous story can be told about the history of polynomial gcd computations. The polynomial counterpart of Euclid's algorithm was first described by Stevin [33, p. 241]. The first algorithms for polynomial half-gcds are due to Moenck [25] and Brent–Gustavson–Yun [5]; several variants have been developed since [35, 36, 4].

We refer to [10, chapter 11] for a gentle exposition of the polynomial half-gcd algorithm. This exposition also contains a careful analysis of the constant factor in the asymptotic time complexity. More precisely, let $\mathsf{M}(d)$ be the complexity to multiply two polynomials of degree $<d$ over an abstract effective field $\mathbb{K}$. Then the gcd of two polynomials of degree $<d$ can be computed using $\lesssim 22\,\mathsf{M}(d)\log_2 d$ operations in $\mathbb{K}$. This complexity further drops to $\lesssim 10\,\mathsf{M}(d)\log_2 d$ if the corresponding Euclidean remainder sequence is "normal" (all quotients in the ordinary Euclidean algorithm are of degree one). The authors declare that they made no particular efforts to optimize these constants 22 and 10; in [10, Research problem 11.11], they ask the question how far these constants can be lowered. This main goal of the present paper is to make progress on this question.

---

[*]. This article has been written using GNU TEXMACS [18].

|              | MCA [10] | General | FFT model | Binary FFT model |
|--------------|----------|---------|-----------|------------------|
| Normal case  | 10       | $7/2$   | 2         | $4/3$            |
| General case | 22       | $17/4$  | $11/4$    | $19/12$          |

**Table 1.** Summary of the constant factors in various cases.

One major motivation for optimizing constant factors in time complexity bounds is the development of faster implementations. Now practical algorithms for multiplying large polynomials usually rely on fast Fourier techniques. We will also investigate optimizations that are only possible when multiplications are done in this way. Indeed, this so-called FFT model allows for various specific optimizations that typically help to reduce constant factors. We refer to [3] for a nice survey of such tricks and mention NTL [31] as a software library that was early to exploit them in a systematic way.

A particularly favorable case is when the ground field $\mathbb{K}$ has primitive $2^t$-th roots of unity for all sufficiently large $t$. We will call this the "binary FFT model". A good example is the finite field $\mathbb{K} = \mathbb{F}_p$ for a prime $p$ of the form $p = s\, 2^t + 1$, where $t$ is large. Such finite fields arise naturally after modular reduction, provided that $p$ can be chosen freely. The binary FFT model allows for additional tricks and is particularly useful for dichotomic algorithms such as the half-gcd. We refer to section 2 for those results that will be required in this paper. Note that large polynomial multiplications over general finite fields are fastest using FFT-based techniques, although one may not necessarily use radix two FFTs; see [14, 15] and references therein for the fastest current algorithms.

For convenience of the reader, we will first describe our new half gcd algorithms in the normal case (see section 4). This will allow us to avoid various technical complications and focus on the new optimizations. The first idea behind our new half gcd is to make the update step after the first recursive call particularly efficient, by using a $2 \times 2$ matrix variant of the middle product algorithm [12]. This leads to an algorithm of time complexity $\lesssim 7/2\, \mathsf{M}(d) \log_2 d$ in general and $\lesssim 2\, \mathsf{M}(d) \log_2 d$ in the FFT model (Proposition 6). The second idea is to combine this with known optimizations for the (binary) FFT model, such as FFT doubling, FFT caching, and Harvey's variant of the middle product [13]. In the binary FFT model, this allows us to compute half gcds in time $\lesssim 4/3\, \mathsf{M}(d) \log d$ (Proposition 7 and Corollary 9).

When dropping the normality assumption, the algorithm becomes more technical, but it turns out that the middle product trick can still be generalized. This is explained in section 5 and leads to the bound $\lesssim 17/4\, \mathsf{M}(d) \log_2 d$ for general gcds and $\lesssim 11/4\, \mathsf{M}(d) \log_2 d$ in the FFT model (Theorem 15). In the binary FFT model, special efforts are needed in order to enforce the use of DFTs of power of two lengths. We prove the bound $\lesssim 19/12\, \mathsf{M}(d) \log_2 d$ in this last case (Theorem 17). See Table 1 for a summary of the new constant factors.

It is well known that polynomial gcds have many applications: fast arithmetic on rational functions, fast reconstruction of rational functions [10, section 5.7], polynomial factorization [7], fast decoding of error-correcting codes [2, 24, 9], sparse interpolation [1, 28], etc. Personally, we were mainly motivated by the last application to sparse interpolation. After the introduction of the tangent Graeffe algorithm [11, 17], gcd computations have often become one of the main practical bottlenecks. For this application, it is typically possible to work modulo a prime of the form $p = s\, 2^t + 1$, which allows us to exploit our optimizations for the binary FFT model.

We made a first implementation of the new algorithm and also programmed the customary extension to the computation of subresultants (see [10, section 11.2] and also [21] for this extension). Work is in progress on HPC implementations of our algorithms and all required subalgorithms.

A few questions remain for future work. It is likely that the ideas in this paper can be applied to integer gcds as well, except for the optimizations that are specific to the binary FFT model. It would also be nice to have counterparts for the "ternary FFT model", which could be used for fields $\mathbb{K}$ of characteristic two (see [17, section 3.4] for an algorithm of this kind in the case of iterated Graeffe transforms). Finally, in view of Bernstein and Yang's recent ideas from [4], as well as Remark 16, one may wonder whether the bounds for the normal case can actually be extended to the general case.

**Acknowledgments.** We are grateful to Michael MONAGAN for triggering this study and suggesting some early ideas.

## 2. PRELIMINARIES

### 2.1. The binary FFT model

The best bounds in this paper will be proved for the so-called "binary FFT model", which requires special assumptions on the ground field $\mathbb{K}$. First of all, this model requires 2 to be invertible in $\mathbb{K}$. Secondly, for any polynomial $P \in \mathbb{K}[x]$ that occurs during computations, we assume that $\mathbb{K}$ has a primitive $n$-th root of unity $\omega_n$ with $n = 2^k > \deg P$. For convenience, we also assume that $\omega_m = \omega_n^{n/m}$ whenever $m$ divides $n$. We measure computational complexity in terms of the number of required field[1] operations in $\mathbb{K}$. Given $P, Q \in \mathbb{K}[x]$ with $Q \neq 0$, we write $P$ quo $Q$ and $P$ rem $Q$ for the quotient and the remainder of the Euclidean division of $P$ by $Q$, respectively.

Let $\mathbb{K}[x]_{<n}$ denote the space of polynomials of degree $<n$. Given $P \in \mathbb{K}[x]_{<n}$ with $n = 2^k$, we define its *discrete Fourier transform* $\mathrm{DFT}_n(P) \in \mathbb{K}^n$ by

$$\mathrm{DFT}_n(P) := (P(\omega_n^0), P(\omega_n^1), \ldots, P(\omega_n^{n-1})).$$

We will write $\mathsf{F}(n)$ for the maximum cost of computing a discrete Fourier transform of order $n$ or the inverse transformation $\mathrm{DFT}_n^{-1} \colon \mathbb{K}^n \to \mathbb{K}[x]_{<n}$. It is well known [8] that $\mathsf{F}(n) = O(n \log n)$. In what follows, we will always assume that $\mathsf{F}(n)/n$, and $\mathsf{F}(n)/(n \log n)$ are non-decreasing (for $n \geqslant 1$ and $n \geqslant 2$, respectively). Therefore, we may just as well take $\mathsf{F}(n) \asymp n \log n$, but it is instructive to keep the notation $\mathsf{F}(n)$ to indicate where we use Fourier transforms.

If $n \geqslant 2$ and the discrete Fourier transform of $P \in \mathbb{K}[x]_{<n}$ at order $n/2$ with respect to $\omega_{n/2}$ is known, then it costs $\mathsf{F}(n/2) + O(n)$ to compute $\mathrm{DFT}_n(P)$. Indeed, we already know $(P(\omega_n^0), P(\omega_n^2), \ldots, P(\omega_n^{n-2}))$, so it remains to compute $(P(\omega_n^1), P(\omega_n^3), \ldots, P(\omega_n^{n-1}))$. But this is the DFT of $\tilde{P} := P(\omega_n x)$ rem $(x^{n/2} - 1)$ at order $n/2$ and it takes a linear number of operations to compute $\tilde{P}$ in terms of $P$. We call this trick *FFT doubling*.

---

1. Using an easy refinement of the analysis, it turns out that the main half gcd algorithms in this paper involve only a linear number of divisions, so the bulk of the operations are actually ring operations in $\mathbb{K}$.

Given $P, Q \in \mathbb{K}[x]$ with $PQ \in \mathbb{K}[x]_{<n}$, we may use the discrete Fourier transform to compute the product $PQ$ using

$$PQ = \mathrm{DFT}_n^{-1}(\mathrm{DFT}_n(P) \, \mathrm{DFT}_n(Q)). \tag{1}$$

This is called *FFT multiplication*. If we fix $P$ of degree $d < n$, then we may consider the multiplication map with $P$ as a linear map $\times_P \colon \mathbb{K}[x]_{<n-d} \to \mathbb{K}[x]_{<n}$. We have

$$\times_P = \mathrm{DFT}_n^{-1} \circ \times_{\mathrm{DFT}_n(P)} \circ \mathrm{DFT}_n \circ \iota_{n-d},$$

where $\times_{\mathrm{DFT}_n(P)}$ stands for componentwise multiplication by $\mathrm{DFT}_n(P)$ in $\mathbb{K}^n$ and $\iota_{n-d}$ for the injection of $\mathbb{K}[x]_{<n-d}$ into $\mathbb{K}[x]_{<n}$.

Given $P \in \mathbb{K}[x]_{<n}$ of degree $d$ and $R \in \mathbb{K}[x]_{<n}$, we define their *middle product* $P \bowtie_d R$ by

$$P \bowtie_d R = \sum_{i=0}^{n-d-1} \left[ \sum_{k=0}^{d} P_k R_{d+i-k} \right] x^i. \tag{2}$$

Let us recall how $P \bowtie_d R$ can be computed efficiently using FFT multiplication.

Let $P \in \mathbb{K}[x]$ be of degree $d < n$, let $Q \in \mathbb{K}[x]_{<n-d}$ and $R \in \mathbb{K}[x]_{<n}$. If $R = PQ$, then we observe that

$$\begin{pmatrix} R_0 \\ \vdots \\ \vdots \\ \vdots \\ \vdots \\ R_{n-1} \end{pmatrix} = \begin{pmatrix} P_0 & & \\ \vdots & \ddots & \\ \vdots & & P_0 \\ P_d & & \vdots \\ & \ddots & \vdots \\ & & P_d \end{pmatrix} \begin{pmatrix} Q_0 \\ \vdots \\ Q_{n-1-d} \end{pmatrix}.$$

The matrix in the middle has $n$ rows and $n-d-1$ columns and we may think of it as the matrix of the map $\times_P$. If $Q = P \bowtie_d R$, then we note that

$$\begin{pmatrix} Q_0 \\ \vdots \\ Q_{n-1-d} \end{pmatrix} = \begin{pmatrix} P_d & \cdots & \cdots & P_0 & & \\ & \ddots & & & \ddots & \\ & & P_d & \cdots & \cdots & P_0 \end{pmatrix} \begin{pmatrix} R_0 \\ \vdots \\ \vdots \\ \vdots \\ \vdots \\ R_{n-1} \end{pmatrix}.$$

In other words $Q = (\times_{\tilde{P}}^{\top})(R)$, where $\tilde{P}(x) = x^d P(1/x)$ and $\times_{\tilde{P}}^{\top} \colon \mathbb{K}[x]_{<n} \to \mathbb{K}[x]_{<n-d}$ stands for the transpose of $\times_{\tilde{P}}$. Since

$$\begin{aligned} \times_{\tilde{P}}^{\top} &= (\mathrm{DFT}_n^{-1} \circ \times_{\mathrm{DFT}_n(\tilde{P})} \circ \mathrm{DFT}_n \circ \iota_{n-d})^{\top} \\ &= \iota_{n-d}^{\top} \circ \mathrm{DFT}_n^{\top} \circ \times_{\mathrm{DFT}_n(\tilde{P})} \circ (\mathrm{DFT}_n^{-1})^{\top} \\ &= \pi_{n-d} \circ \mathrm{DFT}_n \circ \times_{\mathrm{DFT}_n(\tilde{P})} \circ \mathrm{DFT}_n^{-1}, \end{aligned}$$

where $\pi_{n-d} \colon \mathbb{K}[x]_{<n} \to \mathbb{K}[x]_{<n-d}; P \mapsto P \operatorname{rem} x^{n-d}$, it follows that

$$Q = P \bowtie_d R = \mathrm{DFT}_n(\mathrm{DFT}_n(\tilde{P}) \, \mathrm{DFT}_n^{-1}(R)) \operatorname{rem} x^{n-d}.$$

Taking DFTs with respect to $\omega_n^{-1}$ instead of $\omega_n$ and using that $\mathrm{DFT}_{n,\omega_n^{-1}} = n \, \mathrm{DFT}_{n,\omega_n}^{-1}$, we also obtain

$$\begin{aligned} Q &= \mathrm{DFT}_{n,\omega^{-1}}(\mathrm{DFT}_{n,\omega^{-1}}(\tilde{P}) \, \mathrm{DFT}_{n,\omega^{-1}}^{-1}(R)) \operatorname{rem} x^{n-d} \\ &= \mathrm{DFT}_{n,\omega^{-1}}(\mathrm{DFT}_{n,\omega^{-1}}(x^d) \, \mathrm{DFT}_{n,\omega^{-1}}(P(x^{-1})) \, \mathrm{DFT}_{n,\omega^{-1}}^{-1}(R)) \operatorname{rem} x^{n-d} \\ &= \mathrm{DFT}_{n,\omega}^{-1}(\mathrm{DFT}_{n,\omega}(x^{-d}) \, \mathrm{DFT}_{n,\omega}(P) \, \mathrm{DFT}_{n,\omega}(R)) \operatorname{rem} x^{n-d} \\ &= \mathrm{DFT}_n^{-1}(\mathrm{DFT}_n(P) \, \mathrm{DFT}_n(R)) \operatorname{quo} x^d. \tag{3} \end{aligned}$$

This is the alternative formula from [13] that we will use.

From the complexity perspective, let $M(d)$ be the cost to multiply two polynomials in $\mathbb{K}[x]_{<d}$. Using FFT multiplication, we see that $M(d) = 3F(n) + O(n) = 6F(d) + O(d)$ in the binary FFT model. More generally, if $n$ is even, then any two polynomials $P, Q \in \mathbb{K}[x]$ with $\deg PQ < n$ can be multiplied in time $3F(n) + O(n) = M(n/2) + O(n)$ using this method. Similarly, the cost $M^\top(d)$ to compute a middle product (2) with $\deg P = d$ and $n = 2d - 1$ satisfies $M^\top(d) = 6F(d) + O(d) = M(d) + O(d)$. If $n$ is even and $0 \leqslant d = \deg P < n$ is general, then $P \bowtie_d R$ can be computed in time $M^\top(n/2) + O(n) = M(n/2) + O(n)$.

It is important to note that the above results all generalize to the case when the coefficient field $\mathbb{K}$ is replaced by the algebra $\mathbb{K}^{r \times r}$ of $r \times r$ matrices with entries in $\mathbb{K}$. For instance, assume that $P, Q \in \mathbb{K}^{2 \times 2}[x]$ with $\deg PQ < n$. Then (1) allows us to compute $PQ$ using 8 DFTs, 4 inverse DFTs, and $n$ multiplications of $2 \times 2$ matrices in $\mathbb{K}^{2 \times 2}$, for a total cost of $12F(n) + O(n) = 4F(d) + O(d)$.

## 2.2. Other models

Although the main focus in this paper is on the binary FFT model, we will also consider other types of polynomial multiplication. In that case, we will still denote by $M(d)$ and $M^\top(d)$ the complexities of multiplying two polynomials in $\mathbb{K}[x]_{<d}$ and to compute the middle product $P \times_d R$ for $P \in \mathbb{K}[x]$ of degree $d$ and $R \in \mathbb{K}[x]_{<2d-1}$. Moreover, we make the following assumptions on these cost functions:

- We have $M^\top(d) \sim M(d)$.

- The product of any $P, Q \in \mathbb{K}[x]$ with $\deg PQ < 2d$ can be computed in time $\lesssim M(d)$ and similarly for the middle product $P \times_k R$ of $P, R \in \mathbb{K}[x]$ with $\deg P = k < 2d$ and $\deg R < 2d$.

- The functions $M(d)$, $M(d)/d$, and $M(d)/(d \log d)$ are non-decreasing (for $d \geqslant 0$, $d \geqslant 1$, and $d \geqslant 2$, respectively).

We also make the same assumptions for the analogue cost functions $M_{2 \times 2}(d)$ and $M_{2 \times 2}^\top(d)$ when taking coefficients in $\mathbb{K}^{2 \times 2}$ instead of $\mathbb{K}$. In the binary FFT model we have seen that we may take $M(d)$ and $M^\top(d)$ to be of the form $\alpha d \log_2 d + \beta d$ for suitable constants $\alpha, \beta$, after which the above assumptions are satisfied (if $d$ is not a power of two, then one may use the truncated Fourier transform [16], for instance). They are also verified for all other commonly used types of multiplication, such as Karatsuba's and Toom's methods [19, 37], or FFT-based methods for arbitrary radices [30, 6, 15].

In addition, we define $\mu_{2 \times 2}$ to be a constant such that $M_{2 \times 2}(d) \sim M_{2 \times 2}^\top(d) \lesssim \mu_{2 \times 2} M(d) \sim \mu_{2 \times 2} M^\top(d) + O(d)$. We have seen that we may take $\mu_{2 \times 2} = 4$ in the binary FFT model; this generalizes to arbitrary FFT models. Using Strassen's method for $2 \times 2$ matrix multiplication [34], we may always take $\mu_{2 \times 2} \leqslant 7$. From a practical point of view, we usually have $4 \leqslant \mu_{2 \times 2} \leqslant 8$.

Let us examine the constant $\mu_{2 \times 2}$ a bit more closely in the case of Karatsuba multiplication. The complexity of this method satisfies

$$\begin{aligned} M(1) &= \alpha \\ M(2d) &= 3M(d) + \beta d \end{aligned}$$

for certain constants $\alpha$ and $\beta$, which yields $\mathsf{M}(2^k) \sim (\alpha + \beta)\, 3^k$. Similarly, when using Karatsuba's method to multiply $2 \times 2$ matrices, the complexity $\mathsf{M}_{2\times 2}(d)$ satisfies

$$\begin{aligned} \mathsf{M}_{2\times 2}(1) &= 8\alpha \\ \mathsf{M}_{2\times 2}(2d) &= 3\,\mathsf{M}_{2\times 2}(d) + 4\beta d, \end{aligned}$$

which leads to $\mathsf{M}_{2\times 2}(2^k) \sim (8\alpha + 4\beta)\, 3^k$ and

$$\mu_{2\times 2} = \frac{8\alpha + 4\beta}{\alpha + \beta}.$$

This analysis can be generalized to general lengths $d$ and to the case when we only use Karatsuba's method for degrees above a certain threshold. In the latter case, it is usually favorable to choose the threshold over $\mathbb{K}^{2\times 2}$ to be slightly lower than the threshold over $\mathbb{K}$.

## 3. EUCLIDEAN REMAINDER SEQUENCES

### 3.1. Definition and basic properties

Let $P, Q \in \mathbb{K}[x]$ be such that $d := \deg P > \deg Q$. The *Euclidean remainder sequence* $(R_k)_{0 \leqslant k \leqslant \ell}$ is defined by

$$\begin{aligned} R_0 &:= P \\ R_1 &:= Q \\ R_{k+1} &:= R_{k-1} \operatorname{rem} R_k. \end{aligned}$$

The *length* of the sequence is the smallest index $\ell$ for which $R_\ell = 0$. For $1 \leqslant k \leqslant \ell$, we set

$$A_k := \begin{pmatrix} R_{k-1} \\ R_k \end{pmatrix}.$$

We also define the sequence of *Bezout matrices* $(B_k)_{1 \leqslant k < \ell}$ by

$$B_k := \begin{pmatrix} 0 & 1 \\ 1 & -R_{k-1} \operatorname{quo} R_k \end{pmatrix},$$

so that

$$A_{k+1} = B_k A_k,$$

for $1 \leqslant k \leqslant \ell - 1$. We have $\gcd(P, Q) = R_{\ell-1}$.

We regard $B_k$ as a matrix polynomial in $\mathbb{K}^{2\times 2}[x]$ and say that $(R_k)_{0 \leqslant k \leqslant \ell}$ and $(B_k)_{1 \leqslant k < \ell}$ are *normal* if $\deg B_k = 1$ for all $k$. This is the case if and only if $\ell = d+1$ and $\deg R_k = \deg R_{k-1} - 1 = d - k$ for all $k \in \{1, \ldots, d\}$. For $i \leqslant j$, we also define

$$B_{i;j} = B_{j-1} \cdots B_{i+1} B_i,$$

so that

$$A_j = B_{i;j} A_i.$$

(We understand that $B_{i;j} = \operatorname{Id}_2$ if $i = j$.) In particular,

$$\begin{pmatrix} \gcd(P, Q) \\ 0 \end{pmatrix} = A_\ell = B_{1;\ell} A_1 = B_{1;\ell} \begin{pmatrix} P \\ Q \end{pmatrix},$$

so an extended gcd computation essentially boils down to the computation of the matrix product $B_{1;\ell} = B_{\ell-1} \cdots B_1$. In the case of a normal remainder sequence, this is done most efficiently using binary splitting:

$$\begin{aligned} B_{i;i+1} &= B_i \\ B_{i;j} &= B_{h;j} B_{i;h}, \qquad i+2 \leqslant j, \quad h := \left\lfloor \frac{i+j}{2} \right\rfloor. \end{aligned}$$

In essence, this is also how the half-gcd algorithm works.

**LEMMA 1.** *For any $1 \leqslant i < j \leqslant \ell$, we have*

$$\begin{aligned} \deg B_{i;j} &= \deg B_i + \cdots + \deg B_{j-1} \\ \deg R_i &= d - \deg B_{1;i+1} \end{aligned}$$

*In particular, if $(B_k)_{1 \leqslant k < \ell}$ is normal, then $\deg B_{i;j} = j - i$ and $\deg R_i = d - i$.*

**Proof.** Let us show by induction on $j - i$ that

$$\begin{aligned} \deg B_{i;j} = (\deg B_{i;j})_{2,2} &= \deg B_i + \cdots + \deg B_{j-1} \\ (\deg B_{i;j})_{\alpha,\beta} &< \deg B_i + \cdots + \deg B_{j-1} \end{aligned}$$

for any $(\alpha, \beta) \in \{(1,1), (1,2), (2,1)\}$. This is clear if $j = i+1$, so assume that $i + 2 \leqslant j$ and let $h := \lfloor (i+j)/2 \rfloor$. Then

$$(B_{i;j})_{2,2} = (B_{h;j})_{2,2} (B_{i;h})_{2,2} + (B_{h;j})_{2,1} (B_{i;h})_{1,2},$$

so the induction hypothesis yields

$$\begin{aligned} \deg (B_{h;j})_{2,2} (B_{i;h})_{2,2} &= \deg (B_{h;j})_{2,2} + \deg (B_{i;h})_{2,2} = \deg B_i + \cdots + \deg B_{j-1} \\ \deg (B_{h;j})_{2,1} (B_{i;h})_{1,2} &= \deg (B_{h;j})_{2,1} + \deg (B_{i;h})_{1,2} < \deg B_i + \cdots + \deg B_{j-1}, \end{aligned}$$

whence $(\deg B_{i;j})_{2,2} = \deg B_i + \cdots + \deg B_{j-1}$. In a similar way, the induction hypothesis yields $(\deg B_{i;j})_{\alpha,\beta} < \deg B_i + \cdots + \deg B_{j-1}$ for all $(\alpha, \beta) \in \{(1,1), (1,2), (2,1)\}$.

As to the second relation, we note that

$$\deg B_i = \deg R_{i-1} - \deg R_i,$$

whence $\deg B_{1;i+1} = \deg B_1 + \cdots + \deg B_i = \deg R_0 - \deg R_i$ □

**Example 2.** Taking $P = x^4 + 2x^3 - x^2 + 4$ and $Q = x^3 + x^2 - 2x + 5$, we obtain the following normal remainder sequence:

$$\begin{aligned} R_0 &= x^4 + 2x^3 - x^2 + 4 \\ R_1 &= x^3 + x^2 - 2x + 5 & B_1 &= \begin{pmatrix} 0 & 1 \\ 1 & -x-1 \end{pmatrix} & B_{1;1} &= \begin{pmatrix} 1 & 0 \\ 0 & 1 \end{pmatrix} \\ R_2 &= -x^2 - 4x - 1 & B_2 &= \begin{pmatrix} 0 & 1 \\ 1 & x-3 \end{pmatrix} & B_{1;2} &= \begin{pmatrix} 0 & 1 \\ 1 & -x-1 \end{pmatrix} \\ R_3 &= 10x + 8 & B_3 &= \begin{pmatrix} 0 & 1 \\ 1 & \frac{1}{10}x + \frac{8}{25} \end{pmatrix} & B_{1;3} &= \begin{pmatrix} 1 & -x-1 \\ x-3 & -x^2+2x+4 \end{pmatrix} \\ R_4 &= \frac{39}{25} & B_4 &= \begin{pmatrix} 0 & 1 \\ 1 & -\frac{250}{39}x - \frac{200}{39} \end{pmatrix} & B_{1;4} &= \begin{pmatrix} x-3 & -x^2+2x+4 \\ \frac{1}{10}x^2 + \frac{1}{50}x + \frac{1}{25} & \frac{-1}{10}x^3 - \frac{3}{25}x^2 + \frac{1}{25}x + \frac{7}{25} \end{pmatrix} \\ R_5 &= 0 & & & B_{1;5} &= \begin{pmatrix} \frac{1}{10}x^2 + \frac{1}{50}x + \frac{1}{25} & \frac{-1}{10}x^3 - \frac{3}{25}x^2 + \frac{1}{25}x + \frac{7}{25} \\ -\frac{39}{25} R_1 & \frac{39}{25} R_0 \end{pmatrix} \end{aligned}$$

## 3.2. Re-indexation of irnormal Euclidean remainder sequences

In the irnormal case, it is convenient to work with an alternative indexation of remainder sequences for which $\deg R_k^* \leqslant d-k$, as in the normal case. Note that we will not consider abnormal remainder sequences until section 5 below, so the reader may safely skip this subsection until there.

Let us now explain our reindexation in detail. For any $i \in \{1, \ldots, \ell-1\}$, let

$$\kappa(i) := d - \deg R_i$$

We also take

$$\kappa(0) := 0$$
$$\kappa(\ell) := d+1.$$

Then we set

$$R_{\kappa(i)}^* := R_i$$
$$A_{\kappa(i)}^* := A_i$$
$$B_{\kappa(i)}^* := B_i.$$

Moreover, for any $k \in \{\kappa(i)+1, \ldots, \kappa(i+1)-1\}$, we define

$$R_k^* := R_{i+1}$$
$$A_k^* := A_{i+1}$$
$$B_k^* := \mathrm{Id}_2.$$

For $1 \leqslant i \leqslant j < d$, we also define

$$B_{i;j}^* = B_{j-1}^* \cdots B_{i+1}^* B_i^*,$$

so that we still have

$$A_j^* = B_{i;j}^* A_i^*.$$

By construction, for $k \in \{1, \ldots, d\}$, we have

$$\deg R_k^* \leqslant d-k$$
$$\deg B_{1;k+1}^* \leqslant k.$$

As before, we will sometimes write $R_k^*(P, Q)$ instead of $R_k^*$ in order to emphasize the dependence on $P$ and $Q$, and similarly for $B_k^*(P, Q)$, etc. Occasionally, when $d$ is not clear from the context, we also write $R_k^*(P, Q, d)$, $B_k^*(P, Q, d)$, etc.

**Example 3.** Taking $P = 2x^4 + 11x^3 + 3x^2 - 4x + 3$ and $Q = x^3 + 5x^2 - 5x - 2$, we have

$$R_0 = 2x^4 + 11x^3 + 3x^2 - 4x + 3$$

$$R_1 = x^3 + 5x^2 - 5x - 2 \qquad B_1 = \begin{pmatrix} 0 & 1 \\ 1 & -2x-1 \end{pmatrix} \quad B_{1;1} = \begin{pmatrix} 1 & 0 \\ 0 & 1 \end{pmatrix}$$

$$R_2 = x+5 \qquad B_2 = \begin{pmatrix} 0 & 1 \\ 1 & -x^2+5 \end{pmatrix} \quad B_{1;2} = \begin{pmatrix} 0 & 1 \\ 1 & -2x-1 \end{pmatrix}$$

$$R_3 = 3 \qquad B_3 = \begin{pmatrix} 0 & 1 \\ 1 & -\frac{1}{3}x - \frac{5}{3} \end{pmatrix} \quad B_{1;3} = \begin{pmatrix} 1 & -2x-1 \\ -x^2+1 & 2x^3+x^2-2x \end{pmatrix}$$

$$R_4 = 0 \qquad B_{1;4} = \begin{pmatrix} -x^2+1 & 2x^3+x^2-2x \\ \frac{1}{3}R_1 & -\frac{1}{3}R_0 \end{pmatrix}$$

After reindexation $\kappa(0)=0, \kappa(1)=1, \kappa(2)=3, \kappa(3)=4$, and $\kappa(4)=5$, we obtain

$$\begin{aligned}
R_0^* &= 2x^4+11x^3+3x^2-4x+3 \\
R_1^* &= x^3+5x^2-5x-2 & B_1^* &= \begin{pmatrix} 0 & 1 \\ 1 & -2x-1 \end{pmatrix} & B_{1;1}^* &= \begin{pmatrix} 1 & 0 \\ 0 & 1 \end{pmatrix} \\
R_2^* &= x+5 & B_2^* &= \begin{pmatrix} 1 & 0 \\ 0 & 1 \end{pmatrix} & B_{1;2}^* &= \begin{pmatrix} 0 & 1 \\ 1 & -2x-1 \end{pmatrix} \\
R_3^* &= x+5 & B_3^* &= \begin{pmatrix} 0 & 1 \\ 1 & -x^2+5 \end{pmatrix} & B_{1;3}^* &= \begin{pmatrix} 0 & 1 \\ 1 & -2x-1 \end{pmatrix} \\
R_4^* &= 3 & B_4^* &= \begin{pmatrix} 0 & 1 \\ 1 & -\frac{1}{3}x-\frac{5}{3} \end{pmatrix} & B_{1;4}^* &= \begin{pmatrix} 1 & -2x-1 \\ -x^2+1 & 2x^3+x^2-2x \end{pmatrix} \\
R_5^* &= 0 & & & B_{1;5}^* &= \begin{pmatrix} -x^2+1 & 2x^3+x^2-2x \\ \frac{1}{3}R_1 & -\frac{1}{3}R_0 \end{pmatrix}
\end{aligned}$$

## 4. THE NORMAL CASE

### 4.1. Statement of the non-optimized algorithm

Let $P, Q \in \mathbb{K}[x]$ be as in the previous section with remainder sequence $(R_k)_{0 \leqslant k \leqslant \ell}$. We will write $B_k(P,Q)$ for the corresponding $k$-th Bezout matrix $B_k$ in case we wish to make the dependency on $P$ and $Q$ clear. Similarly, we define $B_{i;j}(P,Q) := B_{j-1}(P,Q) \cdots B_i(P,Q)$ and $R_k(P,Q) = B_k(P,Q)_{2,1} P + B_k(P,Q)_{2,2} Q$. Given a polynomial $U \in \mathbb{K}[x]$ and indices $i, j$, it will also be convenient to define

$$\begin{aligned}
U_{i;j} &:= U_i + U_{i+1} x + \cdots + U_{j-1} x^{j-1-i} \\
U_{i;} &:= U_{i;\deg U+1}.
\end{aligned}$$

Here we understand that $U_{i;j} := 0$ whenever $j \leqslant i$.

**LEMMA 4.** *Given $1 \leqslant k < \ell$ such that $\deg B_1 = \cdots = \deg B_k = 1$, we have*

$$\begin{aligned}
B_k(P,Q) &= B_k(P_{d-2k;}, Q_{d-2k;}) \\
B_{1;k+1}(P,Q) &= B_{1;k+1}(P_{d-2k;}, Q_{d-2k;}).
\end{aligned}$$

**Proof.** We have $\deg B_{i;j} = j-i$ for all $1 \leqslant i \leqslant j \leqslant \ell$. For $i=1,\ldots,k$, the relation

$$A_i = B_{1;i} A_1$$

thus shows that the coefficient $(R_i)_\alpha$ of degree $\alpha$ in $R_i$ only depends on coefficients $P_\beta$ and $Q_\beta$ of $P$ and $Q$ with $\beta > \alpha - i$. In particular,

$$(R_i)_{d-k-1;} = R_i(x^{d-2k} P_{d-2k;}, x^{d-2k} Q_{d-2k;})_{d-k-1;} = R_i(P_{d-2k;}, Q_{d-2k;})_{k-1;}$$

and

$$-R_{i-1} \text{ quo } R_i = -(R_{i-1})_{d-k-1;} \text{ quo } (R_i)_{d-k-1;} = -R_{i-1}(P_{d-2k;}, Q_{d-2k;}) \text{ quo } R_i(P_{d-2k;}, Q_{d-2k;}),$$

since $\deg R_{i-1} = d+1-i$ and $\deg R_i = d-i$. This shows that

$$B_i(P,Q) = B_i(P_{d-2k;}, Q_{d-2k;}).$$

By induction on $i$, we also obtain $B_{1;i+1}(P,Q) = B_{1;i+1}(P_{d-2k;}, Q_{d-2k;})$. We conclude by taking $i = k$. $\square$

The lemma leads to the following recursive algorithm for the computation of $B_{1;k+1}(P,Q)$:

---

**Algorithm 1**
**Input:** $P, Q \in \mathbb{K}[x]$ and $k \leqslant d = \deg P$ with $\deg B_i(P,Q) = 1$ for all $1 \leqslant i \leqslant k$
**Output:** $B_{1;k+1}(P,Q)$

1. If $k = 1$, then return $\begin{pmatrix} 0 & 1 \\ 1 & -P_{d-2;} \operatorname{quo} Q_{d-2;} \end{pmatrix}$

2. Let $h := \lceil k/2 \rceil$ and $\tilde{h} := k - h$

3. Recursively compute $M := B_{1;h+1}(P_{d-2h;}, Q_{d-2h})$

4. Compute $\begin{pmatrix} \tilde{P}_{d-h-2\tilde{h};} \\ \tilde{Q}_{d-h-2\tilde{h};} \end{pmatrix}$ with $\begin{pmatrix} \tilde{P} \\ \tilde{Q} \end{pmatrix} = M \begin{pmatrix} P \\ Q \end{pmatrix}$

5. Recursively compute $\tilde{M} := B_{1;\tilde{h}+1}(\tilde{P}_{d-h-2\tilde{h};}, \tilde{Q}_{d-h-2\tilde{h}})$

6. Return $\tilde{M} M$

---

PROPOSITION 5. *Algorithm 1 is correct.*

**Proof.** If $k = 1$, then the result follows from Lemma 4. If $k > 1$, then Lemma 4 implies $M = B_{1;h+1}$, whence $\tilde{P} = R_h$ and $\tilde{Q} = R_{h+1}$. For $1 \leqslant \tilde{\imath} \leqslant \tilde{h}$, we have $\deg B_{\tilde{\imath}}(\tilde{P}, \tilde{Q}) = \deg B_{\tilde{\imath}}(R_h, R_{h+1}) = \deg B_{h+\tilde{\imath}}(P,Q) = 1$. This allows us to apply Lemma 4 once more, and obtain $\tilde{M} = B_{1;\tilde{h}+1}(R_h, R_{h+1}) = B_{h+1;k+1}(P,Q)$. We conclude by noting that $\tilde{M} M = B_{h+1;k+1}(P,Q) B_{1;h+1}(P,Q) = B_{1;k+1}(P,Q)$. $\square$

## 4.2. Exploiting the middle product

Let us now show how to compute $\tilde{P}_{d-h-2\tilde{h};}$ and $\tilde{Q}_{d-h-2\tilde{h};}$ efficiently in step 4 using a middle product of $2 \times 2$ matrix polynomials. In order to simplify the exposition, we assume that $k$ is a power of two; in subsection 4.4 we will consider arbitrary lengths. We first decompose all polynomials into blocks of degree $<h$. More precisely, let

$$\begin{aligned} P_{[i]} &= P_{d-4h+ih;d-3h+ih} \\ Q_{[i]} &= Q_{d-4h+ih;d-3h+ih} \\ \tilde{P}_{[i]} &= \tilde{P}_{d-3h+ih;d-2h+ih} \\ \tilde{Q}_{[i]} &= \tilde{Q}_{d-3h+ih;d-2h+ih}, \end{aligned}$$

so that

$$\begin{aligned} P_{d-4h;d+1-h} &= P_{[0]} + P_{[1]} x^h + P_{[2]} x^{2h} + P_{d-h} x^{3h} \\ Q_{d-4h;d-h} &= Q_{[0]} + Q_{[1]} x^h + Q_{[2]} x^{2h} \end{aligned}$$

and

$$\begin{aligned} \tilde{P}_{d-3h;} &= \tilde{P}_{[0]} + \tilde{P}_{[1]} x^h + \tilde{P}_{d-h} x^{2h} \\ \tilde{Q}_{d-3h;} &= \tilde{Q}_{[0]} + \tilde{Q}_{[1]} x^h. \end{aligned}$$

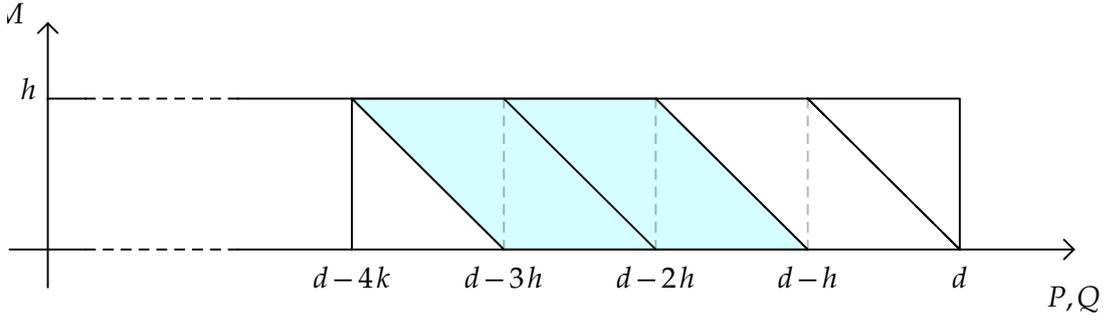

**Figure 1.** Schematic illustration of the computation of $\tilde{P}_{d-3h;d-h}$ and $\tilde{Q}_{d-3h;d-h}$ by taking the middle product of $M$ and the $2 \times 2$ matrix with entries $P_{d-4h;d-2h}$, $P_{d-3h;d-h}$, $Q_{d-4h;d-2h}$, and $Q_{d-3h;d-h}$.

Then we observe that

$$\begin{pmatrix} \tilde{P}_{[0]} & \tilde{P}_{[1]} \\ \tilde{Q}_{[0]} & \tilde{Q}_{[1]} \end{pmatrix} = M \bowtie_h \begin{pmatrix} P_{[0]} + P_{[1]} x^h & P_{[1]} + P_{[2]} x^h \\ Q_{[0]} + Q_{[1]} x^h & Q_{[1]} + Q_{[2]} x^h \end{pmatrix}, \qquad (4)$$

where the left hand matrix has degree $<h$, where $M$ has degree $h$, and where the right hand matrix has degree $<k$; see also Figure 1. The individual term $\tilde{P}_{d-h} x^{2h}$ can be recovered in linear time using

$$\tilde{P}_{d-h} = (M_{1,1} P + M_{1,2} Q)_{d-h} = \sum_{i=0}^{h} [(M_{1,1})_i P_{d-h-i} + (M_{1,2})_i Q_{d-h-i}]. \qquad (5)$$

Before we discuss further optimizations that are specific to the binary FFT model, let us first consider a general multiplication scheme (that satisfies the assumptions from section 2.2), and analyze the complexity of Algorithm 1 with the middle product optimization.

PROPOSITION 6. *The cost of Algorithm 1 with the middle product optimization is bounded by*

$$\frac{\mu_{2\times 2}}{2} \mathsf{M}(k) \log_2 k + O(\mathsf{M}(k)).$$

*Moreover, for a multiplication with $\mathsf{M}(k) \asymp k^\sigma$ for some $1 < \sigma \leqslant 2$, the cost is*

$$\lesssim \frac{\mu_{2\times 2}}{2^{\sigma-1} - 1} \mathsf{M}(k).$$

**Proof.** Recall that we assumed $k$ to be a power of two. Then the running time of the algorithm satisfies the recurrence inequality

$$\mathsf{T}(k) \leqslant 2\mathsf{T}\left(\frac{k}{2}\right) + 2\mu_{2\times 2} \mathsf{M}\left(\frac{k}{2}\right) + O(k). \qquad (6)$$

Unrolling this relation while using the assumption that $\mathsf{M}(k)/(k \log k)$ is non-decreasing yields the first bound. If $\mathsf{M}(k) \sim c k^\sigma$ for some $c > 0$ and $1 < \sigma \leqslant 2$, then the relation (6) yields $\mathsf{T}(k) \lesssim \frac{\mu_{2\times 2} c}{2^{\sigma-1} - 1} k^\sigma$. □

## 4.3. Implementation in the binary FFT model

In order to efficiently implement Algorithm 1 with the optimizations from the previous subsection in the binary FFT model, let us again assume that $k$ is a power of two. We essentially have to compute a $2 \times 2$ matrix middle product in step 4 and a $2 \times 2$ matrix product in step 6. We will do all these computations using DFTs of length $k = 2h$.

Let us first consider the middle product (4). We have $\deg M = h$ in (4) and the right hand side matrix has degree $<k$. Consequently, we may apply (3) and compute the middle product using FFT multiplication:

$$\begin{pmatrix} \tilde{P}_{[0]} & \tilde{P}_{[1]} \\ \tilde{Q}_{[0]} & \tilde{Q}_{[1]} \end{pmatrix} = \mathrm{DFT}_k^{-1}\left(\mathrm{DFT}_k(M)\,\mathrm{DFT}_k\begin{pmatrix} P_{[0]} + P_{[1]}x^h & P_{[1]} + P_{[2]}x^h \\ Q_{[0]} + Q_{[1]}x^h & Q_{[1]} + Q_{[2]}x^h \end{pmatrix}\right) \operatorname{quo} x^h. \quad (7)$$

We also recall that the individual term $\tilde{P}_{d-h} x^{2h}$ can be recovered separately using (5).

As to the matrix product $\tilde{M}M$, its degree is $k$, which is just one too large to use FFT multiplication directly. Nevertheless, FFT multiplication still allows us to compute $\tilde{M}M$ modulo $x^k - 1$. Then we may simply recover $\tilde{M}M$ by computing the leading coefficient $(\tilde{M}M)_k = \tilde{M}_h M_h$ separately.

Now the FFT model also allows for "FFT caching" when doing the recursive calls. In addition to $M\tilde{M}$, we return its DFT transform at the end of the algorithm. When computing the DFT transforms of $M$ and $\tilde{M}$ at length $k$, this means that we already know their DFT transforms at length $h$, and thereby save half of the work.

Summarizing, this leads to the following algorithm:

---

**Algorithm 2**
**Input:** $P, Q \in \mathbb{K}[x]$ and $k \in 2^{\mathbb{N}}$ such that $d = \deg P \geqslant 2k$ and $\deg B_i(P,Q) = 1$ for all $1 \leqslant i \leqslant k$
**Output:** $B_{1;k+1}(P,Q)$ and $\mathrm{DFT}_k(B_{1;k+1}(P,Q))$

1. If $k=1$, then return $B := \begin{pmatrix} 0 & 1 \\ 1 & -P_{d-2;} \operatorname{quo} Q_{d-2;} \end{pmatrix}$ and $\mathrm{DFT}_1(B)$

2. Let $h := k/2$

3. Recursively compute $M := B_{1;h+1}(P_{d-2h;}, Q_{d-2h})$ and $\mathrm{DFT}_h(M)$
   Compute $\hat{M} = \mathrm{DFT}_k(M)$ using FFT doubling

4. Compute $\begin{pmatrix} \tilde{P}_{d-3h;} \\ \tilde{Q}_{d-3h;} \end{pmatrix}$ with $\begin{pmatrix} \tilde{P} \\ \tilde{Q} \end{pmatrix} = M \begin{pmatrix} P \\ Q \end{pmatrix}$ using (7) and (5)

5. Recursively compute $\tilde{M} := B_{1;h+1}(\tilde{P}_{d-3h;}, \tilde{Q}_{d-3h})$ and $\mathrm{DFT}_h(\tilde{M})$
   Compute $\hat{\tilde{M}} = \mathrm{DFT}_k(\tilde{M})$ using FFT doubling

6. Compute $\hat{\tilde{M}}\hat{M}$ and $\tilde{M}M = \mathrm{DFT}_k^{-1}(\hat{\tilde{M}}\hat{M}) + \tilde{M}_h M_h (x^k - 1)$
   Return $\tilde{M}M$ and $\hat{\tilde{M}}\hat{M}$

---

PROPOSITION 7. *Algorithm 2 is correct and its cost is bounded by*

$${}^4\!/_3\, \mathsf{M}(k) \log_2 k + O(\mathsf{M}(k)).$$

**Proof.** Since the algorithm is simply an adaptation of Algorithm 1 to the binary FFT model, it is correct by Proposition 5. Let us analyze the costs of the various steps without the recursive calls.

- The cost of the DFTs in step 3 is bounded by $4\,\mathsf{F}(h) + O(h)$.

- The cost of step 4 is bounded by $8\,\mathsf{F}(k) + O(k)$.

- The cost of the DFTs in step 5 is again bounded by $4\,\mathsf{F}(h) + O(h)$.

- The cost of step 6 is bounded by $4\,\mathsf{F}(k) + O(k)$.

The total cost of the top-level of the algorithm is therefore bounded by $12\,\mathsf{F}(k) + 8\,\mathsf{F}(h) + O(k) = 16\,\mathsf{F}(k) + O(k) = {}^8\!/_3\,\mathsf{M}(k) + O(k)$. Consequently, the cost of our algorithm satisfies the recurrence inequality

$$\mathsf{T}(k) \;\leqslant\; 2\,\mathsf{T}\!\left(\tfrac{k}{2}\right) + \tfrac{8}{3}\,\mathsf{M}(k) + Ck.$$

Unrolling this relation while using the assumption that $\mathsf{M}(k)/(k\log k)$ is non-decreasing yields the desired complexity bound. □

### 4.4. Generalization to arbitrary lengths $k$

Let us now generalize Algorithm 2 to the case when $k$ is not necessarily a power of two.

---

**Algorithm 3**
**Input:** $P, Q \in \mathbb{K}[x]$ and $k \in \mathbb{N}$ such that $d = \deg P \geqslant 2k$ and $\deg B_i(P,Q) = 1$ for all $1 \leqslant i \leqslant k$
**Output:** $B_{1;k+1}(P,Q)$

1. If $k = 0$, then return $\mathrm{Id}_2$
2. Let $h \in 2^{\mathbb{N}}$ be maximal such that $h \leqslant k$ and $\tilde{h} := k - h$
3. Compute $M := B_{1;h+1}(P,Q)$ using Algorithm 2
4. Compute $\begin{pmatrix} \tilde{P}_{d-h-2\tilde{h};} \\ \tilde{Q}_{d-h-2\tilde{h};} \end{pmatrix}$ with $\begin{pmatrix} \tilde{P} \\ \tilde{Q} \end{pmatrix} = M \begin{pmatrix} P \\ Q \end{pmatrix}$
5. Recursively compute $\tilde{M} := B_{1;\tilde{h}+1}(\tilde{P}_{d-h-2\tilde{h};}, \tilde{Q}_{d-h-2\tilde{h}})$
6. Return $\tilde{M} M$

---

PROPOSITION 8. *Algorithm 2 is correct and its cost is bounded by*

$${}^4\!/_3\,\mathsf{M}(k)\log_2 k + O(\mathsf{M}(k)).$$

**Proof.** The correctness is proved in the same way as for Algorithm 1. For some universal constant $C$, the cost $\mathsf{T}(k)$ of the algorithm satisfies the recurrence relation

$$\mathsf{T}(k) \;\leqslant\; {}^4\!/_3\,\mathsf{M}(h)\log_2 h + \mathsf{T}(\tilde{h}) + C\,\mathsf{M}(h).$$

Writing $k = k_1 + \cdots + k_p$ with $k_1,\ldots,k_p \in 2^{\mathbb{N}}$ and $k_1 > \cdots > k_p$, it follows that

$$\begin{aligned}
\mathsf{T}(k) &\leqslant\; \sum_{i=1}^{p} {}^4\!/_3\,\mathsf{M}(k_i)\log_2 k_i + C\sum_{i=1}^{p}\mathsf{M}(k_i) \\
&\leqslant\; \frac{\mathsf{M}(k)}{k}\left(\sum_{i=1}^{p} {}^4\!/_3\,k_i\log_2 k + C\sum_{i=1}^{p} k_i\right) \\
&=\; \mathsf{M}(k)\,({}^4\!/_3 \log_2 k + C),
\end{aligned}$$

where we used our assumption that $\mathsf{M}(k)/k$ is non-decreasing. □

COROLLARY 9. *Let $P, Q \in \mathbb{K}[x]$ and $k \leqslant d$ be such that $\deg P = d$, $\deg Q = d - 1$, and $\deg B_i(P,Q) = 1$ for all $1 \leqslant i \leqslant k$. Then we may compute $B_{1;k+1}(P,Q)$ in time*

$$ {}^4\!/_3\,\mathsf{M}(k)\log_2 k + O(\mathsf{M}(k)).$$

**Proof.** Modulo multiplication of $P$ and $Q$ with $x^k$, we may assume without loss of generality that $d \geqslant 2k$. □

---

**Algorithm 4**
**Input:** $P, Q \in \mathbb{K}[x]$ with $\deg Q < d := \deg P$ and $k \leqslant d$
**Output:** $B^*_{1;k+1}(P, Q)$

---

1. If $Q_{d-k;d} = 0$, then return $\begin{pmatrix} 1 & 0 \\ 0 & 1 \end{pmatrix}$

2. If $k = 1$, then return $\begin{pmatrix} 0 & 1 \\ 1 & -P_{d-2;}\,\mathrm{quo}\,Q_{d-2} \end{pmatrix}$

3. Let $h := \lceil k/2 \rceil$ and $\tilde{h} := k - h$

4. Recursively compute $M := B^*_{1;h+1}(P_{d-2h;}, Q_{d-2h;})$

5. Let $\delta = h - \deg M$, so that $\deg \tilde{P} = d - h + \delta$ and $\deg \tilde{Q} < d - h$
   Compute $\begin{pmatrix} \tilde{P}_{d-h-2\tilde{h}-\delta;} \\ \tilde{Q}_{d-h-2\tilde{h}-\delta;} \end{pmatrix}$ with $\begin{pmatrix} \tilde{P} \\ \tilde{Q} \end{pmatrix} = M \begin{pmatrix} P \\ Q \end{pmatrix}$

6. If $\tilde{Q}_{d-k;} = 0$, then return $M$

7. If $\delta > 0$, then
   - Compute $D := \tilde{P}_{d-h-2\tilde{h}-\delta;}\,\mathrm{quo}\,\tilde{Q}_{d-h-2\tilde{h}-\delta;}$ and $J = \begin{pmatrix} 0 & 1 \\ 1 & -D \end{pmatrix}$
     We claim that $D = \tilde{P}\,\mathrm{quo}\,\tilde{Q}$
   - Update $M := JM$ and let $h' := k - \deg M \leqslant \tilde{h}$
   - Compute $\begin{pmatrix} \tilde{P}_{d-k-h';} \\ \tilde{Q}_{d-k-h';} \end{pmatrix}$ for the updated $\begin{pmatrix} \tilde{P} \\ \tilde{Q} \end{pmatrix} := \begin{pmatrix} \tilde{Q} \\ \tilde{P} - D\tilde{Q} \end{pmatrix}$

8. Recursively compute $\tilde{M} := B^*_{1;h'+1}(\tilde{P}_{d-k-h';}, \tilde{Q}_{d-k-h';})$

9. Return $\tilde{M} M$

---

**Remark 10.** *Instead of Algorithm 2, we may also use Algorithm 1 with the middle product optimization in step 3. In that case, the complexity bound from Proposition 6 generalizes in a similar way to arbitrary lengths.*

## 5. THE GENERAL CASE

Algorithms 1 and 2 generalize to the abnormal case, modulo several technical adjustments. In this section we describe how to do this.

### 5.1. Statement of the non-optimized algorithm

Let us first show how to adapt Algorithm 1. Lemma 4 now becomes:

LEMMA 11. *Given $1 \leqslant k \leqslant d$, we have*

$$\begin{aligned} B^*_k(P, Q) &= B^*_k(P_{d-2k;}, Q_{d-2k;}) \\ B^*_{1;k+1}(P, Q) &= B^*_{1;k+1}(P_{d-2k;}, Q_{d-2k;}). \end{aligned}$$

**Proof.** We have $\deg B_{1;i} \leqslant i - 1$ for $i = 1, \ldots, k$, whence the relation $A^*_i = B^*_{1;i} A^*_1$ shows that the coefficient $(R^*_i)_\alpha$ of degree $\alpha$ in $R^*_i$ only depends on coefficients $P_\beta$ and $Q_\beta$ of $P$ and $Q$ with $\beta > \alpha - i$. We next proceed in a similar way as in the proof of Lemma 4. □

PROPOSITION 12. *Algorithm 4 is correct.*

**Proof.** If $Q_{d-k;d}=0$, then the result is obvious. If $k=1$ and $Q_{d-1}\neq 0$, then the result follows from Lemma 11. Assume from now on that $k>1$ and $Q_{d-k;d}\neq 0$. Then Lemma 11 implies $M=B^*_{1;h+1}$, whence $\tilde{P}=R^*_h$ and $\tilde{Q}=R^*_{h+1}$.

Let $i$ be largest with $\kappa(i)\leqslant h$. If $\tilde{Q}_{d-k;}=0$ in step 6, then $\kappa(i+1)\geqslant k$, so $B^*_{1;k+1}(P,Q)=B^*_{1;h+1}(P,Q)=M$ and our algorithm returns the correct answer. Assume from now on that $\tilde{Q}_{d-k;}\neq 0$.

We call $\delta:=h-\deg M$ the *degeneracy* after $h$ steps. Let $i$ still be largest with $\kappa(i)\leqslant h$. Then we have $M=B^*_{1;\kappa(i)+1}(P,Q)$ and $\deg B^*_{1;\kappa(i)+1}(P,Q)=\kappa(i)=h-\delta$. In particular, we see that $\tilde{P}=R^*_{\kappa(i)}=R_i$, $\tilde{Q}=R^*_{\kappa(i)+1}=R_{i+1}$, and $\deg \tilde{P}=d-h+\delta$. Furthermore, $\tilde{Q}_{d-k;}\neq 0$ implies that $\deg \tilde{Q}\geqslant d-k$, so $\deg \tilde{Q}-(d-h-2\tilde{h}-\delta)=\deg \tilde{Q}-(d-k-\tilde{h}-\delta)\geqslant \tilde{h}+\delta$. Therefore, we computed the $\tilde{h}+\delta+1$ leading terms of $\tilde{Q}$ as part of $\tilde{Q}_{d-h-2\tilde{h}-\delta;}$. Now $\deg \tilde{P}-\deg \tilde{Q}=d-h+\delta-\deg \tilde{Q}\leqslant \tilde{h}+\delta$, so we only need the $\tilde{h}+\delta+1$ leading coefficients of $\tilde{P}$ and $\tilde{Q}$ in order to compute $\tilde{P}$ quo $\tilde{Q}$. This proves our claim that $D=\tilde{P}$ quo $\tilde{Q}$.

After step 7, we thus have $\tilde{P}=R_{i+1}$, $\tilde{Q}=R_{i+2}$, and $M=B^*_{1;\kappa(i+1)+1}(P,Q)$, where $\deg M=\kappa(i+1)$. Moreover, $\deg R_{i+1}=d-\deg M=d-k+h'$ and $\deg D=d-h+\delta-(d-k+h')=\tilde{h}-h'+\delta$. In particular, we may indeed retrieve the new value of $\tilde{Q}_{d-k-h';}$ from $D$ and the old values of $\tilde{P}_{d-h-2\tilde{h}-\delta;}$ and $\tilde{Q}_{d-h-2\tilde{h}-\delta;}$ at the end of step 7, since $d-k-h'-(d-h-2\tilde{h}-\delta)=\tilde{h}-h'+\delta=\deg D$. Furthermore, $\deg R_{i+1}-(d-k-h')=2h'$, which allows to apply Lemma 11, and obtain $\tilde{M}=B^*_{1;h'+1}(R_{i+1},R_{i+2})$. We conclude that

$$\begin{aligned}\tilde{M}M &= B^*_{1;h'+1}(R_{i+1},R_{i+2})\, B^*_{1;\kappa(i+1)+1}(P,Q)\\ &= B^*_{\kappa(i+1)+1;\kappa(i+1)+h'+1}(P,Q)\, B^*_{1;\kappa(i+1)+1}(P,Q)\\ &= B^*_{1;\kappa(i+1)+h'+1}(P,Q)\\ &= B^*_{1;k+1}(P,Q)\end{aligned}$$

□

**Remark 13.** For efficiency, we chose $h=\lceil k/2\rceil$ in step 3, but it is interesting to note that the above correctness proof actually works for any choice of $h$ with $0<h<k$. We will use this property for our FFT version in section 5.3 below, where we will take $h$ to be a power of two.

## 5.2. Exploiting the middle product

Contrary to what we did in section 4.2, we do not require $k$ to be a power of two in this subsection. In fact, it is possible to efficiently implement step 5 in general, using middle products. This time, we break up our input and output polynomials as follows:

$$\begin{aligned}P_{[i,i+1]} &:= P_{d-2k+i(h+\delta);d-k+i(h+\delta)}\\ Q_{[i,i+1]} &:= Q_{d-2k+i(h+\delta);d-k+i(h+\delta)}\\ \tilde{P}_{[i]} &:= \tilde{P}_{d-k+(i-1)(h+\delta);d-k+i(h+\delta)}\\ \tilde{Q}_{[i]} &:= \tilde{Q}_{d-k+(i-1)(h+\delta);d-k+i(h+\delta)}.\end{aligned}$$

Then we have (see See Figure 1)

$$\begin{pmatrix}\tilde{P}_{[0]} & \tilde{P}_{[1]}\\ \tilde{Q}_{[0]} & \tilde{Q}_{[1]}\end{pmatrix} = M \bowtie_{h-\delta} \begin{pmatrix}P_{[0,1]} & P_{[1,2]}\\ Q_{[0,1]} & Q_{[1,2]}\end{pmatrix}. \tag{8}$$

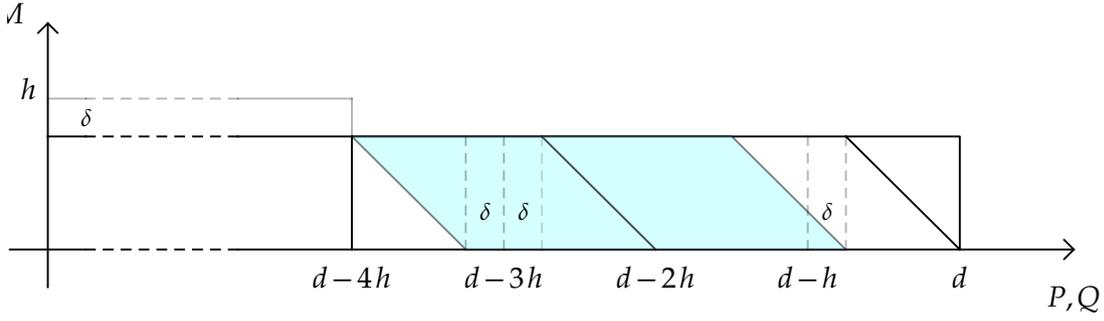

**Figure 2.** Schematic illustration of the matricial middle product in the abnormal case, for even $k = 2h$.

As before, the coefficient $\tilde{P}_{d-h+\delta}$ is computed separately using

$$\tilde{P}_{d-h+\delta} = \sum_{i=0}^{h-\delta} [(M_{1,1})_i P_{d-h+\delta-i} + (M_{1,2})_i Q_{d-h+\delta-i}]. \tag{9}$$

**Remark 14.** In the continuation of Remark 13, we note that the above formulas again apply for any choice of $h$ with $0 < h < k$.

Let $\mu_{2\times 2}$ be as in section 4.2. For the multiplication method that we selected, assume that the middle product (8) and the final product $\tilde{M}M$ can each be computed in time $\leqslant 1/2\, \mu_{2\times 2} \mathsf{M}(k) + O(k)$. Then we have the following generalization of Proposition 6:

THEOREM 15. *Let $4 \leqslant \mu_{2\times 2} \leqslant 8$ be as in Proposition 6. Then Algorithm 4 with the middle product optimization runs in time at most*

$$\frac{2\mu_{2\times 2}+3}{4} \mathsf{M}(k) \log_2 k + O(k \log_2 k).$$

*Moreover, for a multiplication with $\mathsf{M}(k) \asymp k^\sigma$ for some $1 < \sigma \leqslant 2$, the cost is*

$$\lesssim \frac{2\mu_{2\times 2}+3}{2^\sigma - 2} \mathsf{M}(k).$$

**Proof.** The cost of steps 5 and 9 is bounded by $\leqslant \mu_{2\times 2} \mathsf{M}(k) + O(k)$, according to the assumption just above this proposition. In step 7, the computation of $D$ can be done in time $A\,\mathsf{M}(\delta) + O(1)$ with $A \leqslant 4$, using Newton's method for fast power series inversion; see, e.g. [3, section 6]. The update of $M$ amounts to two products of degrees $<h$ by $<\delta$, which can certainly be computed in time $\lesssim 2\,\mathsf{M}(h)$. The update of $\tilde{Q}_{d-k-h'}$ can be computed efficiently using a middle product that takes $\lesssim \mathsf{M}(h)$ additional operations. Altogether, the cost $\mathsf{T}(k)$ of the algorithm satisfies

$$\mathsf{T}(k) \leqslant \mathsf{T}(h) + \mathsf{T}(h') + \left(\mu_{2\times 2} + \tfrac{3}{2}\right) \mathsf{M}(k) + 4\,\mathsf{M}(\delta) + Ck,$$

for some constant $C \geqslant 0$. Note also that $h + h' + \delta \leqslant k$ and that

$$\varphi(k) := \frac{\mathsf{M}(k)}{\max(k,1) \log_2 \max(k,2)}$$

is a non-decreasing function. Let $\alpha := \frac{\mu_{2\times2}}{2} + 1$ and $\beta \geqslant \frac{2C}{\varphi(1)} + 3\alpha$ be such that

$$\mathsf{T}(k) \;\leqslant\; \varphi(k)\,(\alpha \log_2 k + \beta)\,k \max\,(\log_2 k, 1) \tag{10}$$

for all $k \leqslant 2^{\log_2(A/\alpha)+2}$. Let us show by induction on $k$ that (10) holds for all $k > 2^{\log_2(A/\alpha)+2}$ as well. Indeed,

$$\begin{aligned}
\mathsf{T}(k) \;&\leqslant\; \varphi(k)\,(\alpha \log_2^2 h + \beta \log_2 h)\,(h+h') \\
&\quad + 2\alpha\,\varphi(k)\,k \log_2 k + A\,\varphi(k)\,\delta \log_2 k + Ck \\
&\leqslant\; \alpha\,\varphi(k)\,(\log_2^2 k - 2\log_2 k + {}^3\!/\!_2)\,(h+h') + \beta\,\varphi(k)(\log_2 k - {}^1\!/\!_2)\,(h+h') \\
&\quad + 2\alpha\,\varphi(k)\,k \log_2 k + A\,\varphi(k)\,\delta \log_2 k + Ck \\
&\leqslant\; \alpha\,\varphi(k)\,(\log_2^2 k - 2\log_2 k + {}^3\!/\!_2)\,(k-\delta) + 2\alpha\,\varphi(k)\,k \log_2 k + A\,\varphi(k)\,\delta \log_2 k \\
&\quad + \beta\,\varphi(k)\,(\log_2 k - {}^1\!/\!_2)\,k + Ck \\
&\leqslant\; \alpha\,\varphi(k)\,k \log_2^2 k - \alpha\,\varphi(k)\,(\log_2^2 k - 2\log_2 k + {}^3\!/\!_2)\,\delta + A\,\varphi(k)\,\delta \log_2 k \\
&\quad + \beta\,\varphi(k)\,k \log_2 k + (^C\!/\!_{\varphi(1)} + {}^3\!/\!_2\alpha - {}^1\!/\!_2\beta)\,\varphi(k)\,k \\
&\leqslant\; \varphi(k)\,(\alpha \log_2 k + \beta)\,k \log_2 k - \varphi(k)\,(\alpha \log_2^2 k - (A+2\alpha) \log_2 k + {}^3\!/\!_2\alpha)\,\delta \\
&\leqslant\; \varphi(k)\,(\alpha \log_2 k + \beta)\,k \log_2 k.
\end{aligned}$$

This completes the proof of the first bound. We skip the proof of the second one, which is based on similar arguments. □

**Remark 16.** The constants in the bounds from Theorem 15 are probably not sharp. The point is the following: if the quotients in the Euclidean remainder sequence are all of bounded degree, then the additional cost with respect to the normal case can actually be bounded by $O(\mathsf{M}(k))$. If, on the contrary, some of the quotients have exceptionally large degrees, then many of the recursive calls will terminate early at step 6. This will actually make the constants *decrease* instead of increase. The only problematic case therefore seems to be when many of the quotients have moderately large degrees; it would be interesting to know more about the precise worse case scenario.

### 5.3. Implementation in the binary FFT model

One important feature of Algorithm 2 is that we use the same length $k$ for all our DFTs. If $k$ is a power of two, then we may conserve this property in the abnormal case. Indeed, the final product $\tilde{M}M$ causes no problem since its degree is still $<k$. As to the middle product (8), we now have $\deg M = h - \delta$ and the degree of the right hand side is still $<k$. This allows us to apply (3), which yields

$$\begin{pmatrix} \tilde{P}_{[0]} & \tilde{P}_{[1]} \\ \tilde{Q}_{[0]} & \tilde{Q}_{[1]} \end{pmatrix} \;=\; \mathrm{DFT}_k^{-1}\!\left(\mathrm{DFT}_k(M)\,\mathrm{DFT}_k\!\begin{pmatrix} P_{[0,1]} & P_{[1,2]} \\ Q_{[0,1]} & Q_{[1,2]} \end{pmatrix}\right) \mathrm{quo}\; x^{h-\delta}. \tag{11}$$

However, there is no reason why $h'$ should be a power of two for the second recursive call. In order to remedy to this problem, we introduce a new parameter $\ell \geqslant k$ that we assume to be a power of two and that we will use for the lengths of our DFTs. This requires a minor adjustment of (11):

$$\begin{pmatrix} \tilde{P}_{[0]} & \tilde{P}_{[1]} \\ \tilde{Q}_{[0]} & \tilde{Q}_{[1]} \end{pmatrix} \;=\; \mathrm{DFT}_\ell^{-1}\!\left(\mathrm{DFT}_\ell(M)\,\mathrm{DFT}_\ell\!\begin{pmatrix} P_{[0,1]} & P_{[1,2]} \\ Q_{[0,1]} & Q_{[1,2]} \end{pmatrix}\right) \mathrm{quo}\; x^{h-\delta}. \tag{12}$$

We are now in a position to adapt Algorithm 4 to the binary FFT model: see Algorithm 5.

---

**Algorithm 5**
**Input:** $P, Q \in \mathbb{K}[x]$ with $\deg Q < \deg P$ and $k \leqslant \ell \in 2^\mathbb{N}$ with $k \leqslant \deg P$
**Output:** $B^*_{1;k+1}(P,Q)$ and $\mathrm{DFT}_\ell(B^*_{1;k+1}(P,Q))$

1. If $Q_{d-k;d} = 0$, then return $I := \begin{pmatrix} 1 & 0 \\ 0 & 1 \end{pmatrix}$ and $\mathrm{DFT}_\ell(I)$

2. If $\ell = 1$, then return $B := \begin{pmatrix} 0 & 1 \\ 1 & -P_{d-2;}\,\mathrm{quo}\,Q_{d-2;} \end{pmatrix}$ and $\mathrm{DFT}_1(B)$

3. If $k \leqslant \ell/2$, then
   - Recursively compute $M := B^*_{1;k+1}(P,Q)$ and $\mathrm{DFT}_{\ell/2}(M)$
   - Return $M$ and $\mathrm{DFT}_\ell(M)$, which we compute using FFT doubling

4. Let $h := \ell/2$

5. Recursively compute $M := B^*_{1;h+1}(P_{d-2h;},Q_{d-2h;})$ and $\mathrm{DFT}_{\ell/2}(M)$
   Compute $\hat{M} = \mathrm{DFT}_\ell(M)$ using FFT doubling

6. Let $\delta = h - \deg M$, so that $\deg \tilde{P} = d - h + \delta$ and $\deg \tilde{Q} < d - h$
   Compute $\begin{pmatrix} \tilde{P}_{d-3h-\delta;} \\ \tilde{Q}_{d-3h-\delta;} \end{pmatrix}$ with $\begin{pmatrix} \tilde{P} \\ \tilde{Q} \end{pmatrix} = M \begin{pmatrix} P \\ Q \end{pmatrix}$ using (9) and (12)

7. If $\tilde{Q}_{d-2h;} = 0$, then return $M$ and $\hat{M}$

8. If $\delta > 0$, then
   - Compute $D := \tilde{P}_{d-h-2\tilde{h}-\delta;}\,\mathrm{quo}\,\tilde{Q}_{d-h-2\tilde{h}-\delta;} = \tilde{P}\,\mathrm{quo}\,\tilde{Q}$ and $J = \begin{pmatrix} 0 & 1 \\ 1 & -D \end{pmatrix}$
   - Compute $\hat{D} := \mathrm{DFT}_\ell(D)$ and deduce $\hat{J} := \mathrm{DFT}_\ell(J)$
   - Update $\hat{M} := \hat{J}\hat{M}$ and $M_{k-h'} := J_{\deg J} M_{h-\delta}$, where $h' := \tilde{h} + \delta - \deg J$
   - Compute $\begin{pmatrix} \tilde{P}_{d-k-h';} \\ \tilde{Q}_{d-k-h';} \end{pmatrix}$ for the updated $\begin{pmatrix} \tilde{P} \\ \tilde{Q} \end{pmatrix} := \begin{pmatrix} \tilde{Q} \\ \tilde{P} - D\tilde{Q} \end{pmatrix}$

9. Recursively compute $\tilde{M} := B^*_{1;h'+1}(\tilde{P}_{d-k-h';},\tilde{Q}_{d-k-h';})$ and $\mathrm{DFT}_{\ell/2}(\tilde{M})$
   Compute $\hat{\tilde{M}} = \mathrm{DFT}_\ell(\tilde{M})$ using FFT doubling

10. Compute $\hat{\tilde{M}}\hat{M}$ and $\tilde{M}M = \mathrm{DFT}_\ell^{-1}(\hat{\tilde{M}}\hat{M}) + \tilde{M}_{h'} M_{k-h'}(x^\ell - 1)$
    Return $\tilde{M}M$ and $\hat{\tilde{M}}\hat{M}$

---

For the complexity analysis, it will be convenient to assume that $\mathsf{M}(d)$ satisfies the properties from section 2.2. In particular, the assumption that $\mathsf{M}(d)/d$ is non-decreasing implies that $\mathsf{M}(d) + \mathsf{M}(d') \leqslant \mathsf{M}(d + d')$ for all $d, d'$. Conversely, in the binary FFT model, it will be convenient to also assume that $\mathsf{M}(d + d') \leqslant \mathsf{M}(d) + \mathsf{M}(d' - 1) + \Lambda(d + d')$ for all $d, d' \geqslant 1$, and some fixed constant $\Lambda$.

**THEOREM 17.** *Algorithm 5 is correct. Moreover, if $\ell \leqslant 2k$, then its cost is bounded by*

$${}^{19}\!/_{12}\,\mathsf{M}(k)\log_2 k + O(\mathsf{M}(k)).$$

**Proof.** The correctness is proved in a similar way as the correctness of Proposition 12, while using Remarks 13 and 14.

The total cost of steps 5, 6, 9, and 10 is $16\,\mathsf{F}(\ell) + O(\ell)$, as in the proof of Proposition 7. As to the update step 8, the computation of $D$ requires $O(\mathsf{M}(\delta))$ operations. The computation of $\hat{D}$ and $\hat{J}$ costs $\sim \mathsf{F}(\delta)\,k/\delta \leqslant \mathsf{F}(\ell) + O(\ell)$, whereas the multiplication $\hat{J}\hat{M}$ takes linear time.

Now let $p := \deg \tilde{P}$ and $q := \tilde{Q}$ before the updates, so that $p = q + \delta + \eta$ with $\eta \geqslant 0$ and $\deg D = p - q$. Since $h' = h - \eta$, the lowest $\eta$ coefficients of $\tilde{Q}_{d-3h;}$ do not matter in step 9. During the update, this means that we essentially need to compute $(D\tilde{Q}_{d-3h-\delta;})_{d-3h+\eta;}$. Now, using FFT multiplication, the computation of

$$(D\tilde{Q}_{d-3h-\delta;})_{d-3h+\delta;} = D \bowtie_{q-p} \tilde{Q}_{d-3h-\eta;d-h-\eta}$$

takes one direct and one inverse DFT of length $\ell$ of total cost $2\,\mathsf{F}(\ell) + O(\ell)$, since we already know $\hat{D}$. The remaining coefficients $(D\tilde{Q}_{d-3h-\delta;})_{d-3h+\eta;d-3h+\delta}$ can be computed in time $O(\mathsf{M}(\delta))$.

If $k > \ell/2$, then $h = \ell/2$ and the above analysis shows that the time complexity $\mathsf{T}(k,\ell)$ of the algorithm satisfies

$$\mathsf{T}(k,\ell) \;\leqslant\; \max_{\ell/2+h'+\delta\leqslant k} \Big(\mathsf{T}(\ell/2,\ell/2) + \mathsf{T}(h',\ell/2) + \tfrac{19}{6}\mathsf{M}(\ell) + A\,\mathsf{M}(\delta) + C\,\ell\Big),$$

for suitable constants $A$ and $C$. If $k \leqslant \ell/2$, then the FFT doubling in step 3 can be done in time $2\,\mathsf{F}(\ell) + O(\ell)$, so

$$\mathsf{T}(k,\ell) \;\leqslant\; \mathsf{T}(k,\ell/2) + \tfrac{1}{3}\mathsf{M}(\ell) + C\,\ell,$$

by increasing $C$ if necessary. In the bound for $\mathsf{T}(k,\ell)$ when $k > \ell/2$, the term $A\,\mathsf{M}(\delta)$ pollutes the complexity analysis. Our next objective is to reduce to the case when the sum $\mathsf{T}(h',\ell/2) + A\,\mathsf{M}(\delta)$ is replaced by $\mathsf{T}(k-\ell/2,\ell)$.

We start with the definition of an upper bound $\bar{\mathsf{T}}(k,\ell)$ for $\mathsf{T}(k,\ell)$ as follows. For $k \leqslant 1$, we take $\bar{\mathsf{T}}(k,1) := \mathsf{T}(k,1)$. For $\ell/2 < k \leqslant \ell \in 2^{\mathbb{N}}$ with $\ell \geqslant 2$, we define

$$\bar{\mathsf{T}}(k,\ell) \;:=\; \max_{\ell/2+h'+\delta\leqslant k} \Big(\bar{\mathsf{T}}(\ell/2,\ell/2) + \bar{\mathsf{T}}(h',\ell/2) + \tfrac{19}{6}\mathsf{M}(\ell) + A\,\mathsf{M}(\delta) + C\,\ell\Big).$$

For $k \leqslant \ell/2$ with $2 \leqslant \ell \in 2^{\mathbb{N}}$, we take

$$\bar{\mathsf{T}}(k,\ell) \;:=\; \bar{\mathsf{T}}(k,\ell/2) + \tfrac{1}{3}\mathsf{M}(\ell) + C\,\ell. \tag{13}$$

Using an easy induction, we note that $\bar{\mathsf{T}}(k,\ell)$ is increasing in $k$ for fixed $\ell$. If $\ell \geqslant 2$ and $\ell/2 < k \leqslant \ell$, then there exist $h'$ and $\delta$ with $\ell/2 + h' + \delta \leqslant k$ such that

$$\bar{\mathsf{T}}(k,\ell) \;=\; \bar{\mathsf{T}}(\ell/2,\ell/2) + \bar{\mathsf{T}}(h',\ell/2) + \tfrac{19}{6}\mathsf{M}(\ell) + A\,\mathsf{M}(\delta) + C\,\ell. \tag{14}$$

Given $\delta'$ with $k' := k + \delta' \leqslant \ell$, it follows that

$$\begin{aligned}
\bar{\mathsf{T}}(k',\ell) &\geqslant \bar{\mathsf{T}}(\ell/2,\ell/2) + \bar{\mathsf{T}}(h',\ell/2) + \tfrac{19}{6}\mathsf{M}(\ell) + A\,\mathsf{M}(\delta+\delta') + C\,\ell \\
&\geqslant \bar{\mathsf{T}}(\ell/2,\ell/2) + \bar{\mathsf{T}}(h',\ell/2) + \tfrac{19}{6}\mathsf{M}(\ell) + A\,\mathsf{M}(\delta) + A\,\mathsf{M}(\delta') + C\,\ell \\
&= \bar{\mathsf{T}}(k,\ell) + A\,\mathsf{M}(\delta).
\end{aligned}$$

More generally, for any $0 \leqslant k < k' \leqslant \ell$, we claim that

$$\bar{\mathsf{T}}(k,\ell) + A\,\mathsf{M}(k'-k) \;\leqslant\; \bar{\mathsf{T}}(k',\ell) + 2\Lambda\,\ell,$$

where $\Lambda$ is the constant from before the statement of this theorem.

We prove our claim by induction on the smallest $i$ with $k > \ell/2^i$. We already dealt with the case when $i = 1$, so assume that $i > 1$. If $k' \leqslant \ell/2$, then (13) and the induction hypothesis with $\ell/2$ in the role of $\ell$ yield

$$\begin{aligned}
\bar{\mathsf{T}}(k',\ell) - \bar{\mathsf{T}}(k,\ell) &= \bar{\mathsf{T}}(k',\ell/2) - \bar{\mathsf{T}}(k,\ell/2) \\
&\geqslant A\,\mathsf{M}(k'-k) - \Lambda\,\ell.
\end{aligned}$$

In particular,
$$\bar{\mathsf{T}}(k,\ell) + A\,\mathsf{M}(\ell/2 - k) \leqslant \bar{\mathsf{T}}(\ell/2,\ell) + \Lambda\ell.$$

If $k' > \ell/2$, then we have shown above (with $\ell/2$ in the role of $k$) that
$$\bar{\mathsf{T}}(\ell/2 + 1, \ell) + A\,\mathsf{M}(k' - (\ell/2 + 1)) \leqslant \bar{\mathsf{T}}(k', \ell),$$

whence
$$\begin{aligned}
\bar{\mathsf{T}}(k,\ell) + A\,\mathsf{M}(k' - k) &\leqslant \bar{\mathsf{T}}(k,\ell) + A\,\mathsf{M}(\ell/2 - k) + A\,\mathsf{M}(k' - (\ell/2 + 1)) + \Lambda k' \\
&\leqslant \bar{\mathsf{T}}(\ell/2,\ell) + A\,\mathsf{M}(k' - (\ell/2 + 1)) + \Lambda\ell + \Lambda k' \\
&\leqslant \bar{\mathsf{T}}(\ell/2 + 1, \ell) + A\,\mathsf{M}(k' - (\ell/2 + 1)) + \Lambda\ell + \Lambda k' \\
&\leqslant \bar{\mathsf{T}}(k', \ell) + 2\Lambda\ell,
\end{aligned}$$

as claimed.

Now consider $\ell/2 < k \leqslant \ell \in 2^{\mathbb{N}}$ with $\ell \geqslant 2$ and let $h'$ and $\delta$ be such that $\ell/2 + h' + \delta \leqslant k$. Then our claim implies
$$\bar{\mathsf{T}}(h', \ell/2) + A\,\mathsf{M}(\delta) \leqslant \bar{\mathsf{T}}(h' + \delta, \ell/2) + \Lambda\ell \leqslant \bar{\mathsf{T}}(h - \ell/2, \ell/2) + \Lambda\ell.$$

Plugging this into (14), while setting $C' := C + \Lambda$, we obtain
$$\bar{\mathsf{T}}(k,\ell) \leqslant \bar{\mathsf{T}}(\ell/2, \ell/2) + \bar{\mathsf{T}}(k - \ell/2, \ell/2) + \tfrac{19}{6}\mathsf{M}(\ell) + C'\ell. \tag{15}$$

for all $\ell \geqslant 2$ and $\ell/2 < k \leqslant \ell$. Unrolling this inequality for $k = \ell$, we deduce that there exists a constant $C''$ with
$$\bar{\mathsf{T}}(\ell,\ell) \leqslant \tfrac{19}{12}\mathsf{M}(\ell)\log_2 \ell + C''\mathsf{M}(\ell)$$

for all $\ell \in 2^{\mathbb{N}}$. For general $k = k_1 + \cdots + k_p$ with $k_1, \ldots, k_p \in 2^{\mathbb{N}}$ and $k_1 > \cdots > k_p$, combining this bound with (13) and (15) yields
$$\begin{aligned}
\bar{\mathsf{T}}(k,\ell) &\leqslant \frac{19}{12}\sum_{i=1}^{p}\mathsf{M}(k_i)\log_2 k_i + C''\sum_{i=1}^{p}\mathsf{M}(k_i) + \frac{19}{6}\sum_{i=1}^{\log_2\ell}\mathsf{M}(2^i) + C'\sum_{i=1}^{\log_2\ell} 2^i + O(1) \\
&\leqslant \frac{\mathsf{M}(k)}{k}\left(\frac{19}{12}\sum_{i=1}^{p}k_i\log_2 k + C''\sum_{i=1}^{p}k_i\right) + \frac{19}{3}\mathsf{M}(\ell) + 2C'\ell + O(1) \\
&= \mathsf{M}(k)\left(\tfrac{19}{12}\log_2 k + C''\right) + \tfrac{19}{3}\mathsf{M}(\ell) + 2C'\ell + O(1).
\end{aligned}$$

Under the assumption that $\ell \leqslant 2k$, we have $\tfrac{19}{3}\mathsf{M}(\ell) + 2C'\ell = O(\mathsf{M}(k))$, whence $\mathsf{T}(k,\ell) \leqslant \bar{\mathsf{T}}(k,\ell) \leqslant \tfrac{19}{12}\mathsf{M}(k)\log_2 k + O(\mathsf{M}(k))$. □

## BIBLIOGRAPHY

[1] M. Ben-Or and P. Tiwari. A deterministic algorithm for sparse multivariate polynomial interpolation. In *Proc. ACM STOC '88*, pages 301–309. New York, NY, USA, 1988.

[2] E. R. Berlekamp. *Algebraic coding theory*. McGraw-Hill, 1968.

[3] D. J. Bernstein. *Fast multiplication and its applications*, pages 325–384. Mathematical Sciences Research Institute Publications. Cambridge University Press, United Kingdom, 2008.

[4] D. J. Bernstein and B.-Y. Yang. Fast constant-time gcd computation and modular inversion. *IACR Trans. Cryptogr. Hardw. Embed. Syst.*, 3:340–398, 2019.

[5] R. P. Brent, F. G. Gustavson, and D. Y. Y. Yun. Fast solution of Toeplitz systems of equations and computation of Padé approximants. *J. Algorithms*, 1(3):259–295, 1980.

[6] D. G. Cantor and E. Kaltofen. On fast multiplication of polynomials over arbitrary algebras. *Acta Informatica*, 28:693–701, 1991.

[7] D. G. Cantor and H. Zassenhaus. A new algorithm for factoring polynomials over finite fields. *Math. Comp.*, 36(154):587–592, 1981.

[8] J. W. Cooley and J. W. Tukey. An algorithm for the machine calculation of complex Fourier series. *Math. Computat.*, 19:297–301, 1965.

[9] J. L. Dornstetter. On the equivalence between Berlekamp's and Euclid's algorithms. *IEEE Transactions on Information Theory*, 33:428–431, 1987.

[10] J. von zur Gathen and J. Gerhard. *Modern Computer Algebra*. Cambridge University Press, New York, NY, USA, 3rd edition, 2013.

[11] B. Grenet, J. van der Hoeven, and G. Lecerf. Randomized root finding over finite fields using tangent Graeffe transforms. In *Proc. ISSAC '15*, pages 197–204. New York, NY, USA, 2015. ACM.

[12] G. Hanrot, M. Quercia, and P. Zimmermann. The middle product algorithm I. speeding up the division and square root of power series. *AAECC*, 14:415–438, 2004.

[13] D. Harvey. Faster algorithms for the square root and reciprocal of power series. *Math. Comp.*, 80:387–394, 2011.

[14] D. Harvey and J. van der Hoeven. Faster polynomial multiplication over finite fields using cyclotomic coefficient rings. *J. of Complexity*, 54, 2019. Article ID 101404, 18 pages.

[15] D. Harvey and J. van der Hoeven. Polynomial multiplication over finite fields in time $O(n \log n)$. Technical Report, HAL, 2019. http://hal.archives-ouvertes.fr/hal-02070816.

[16] J. van der Hoeven. The truncated Fourier transform and applications. In *Proc. ISSAC 2004*, pages 290–296. Univ. of Cantabria, Santander, Spain, July 2004.

[17] J. van der Hoeven and M. Monagan. Computing one billion roots using the tangent Graeffe method. *ACM SIGSAM Commun. Comput. Algebra*, 54(3):65–85, 2021.

[18] J. van der Hoeven et al. GNU TeXmacs. https://www.texmacs.org, 1998.

[19] A. Karatsuba and J. Ofman. Multiplication of multidigit numbers on automata. *Soviet Physics Doklady*, 7:595–596, 1963.

[20] D. E. Knuth. The analysis of algorithms. In *Actes du congrès international des matheématiciens 1970*, volume 3, pages 269–274. Gauthier-Villars, 1971.

[21] G. Lecerf. On the complexity of the Lickteig–Roy subresultant algorithm. *J. Symbolic Comput.*, 2018. https://doi.org/10.1016/j.jsc.2018.04.017.

[22] D. H. Lehmer. Euclid's algorithm for large numbers. *Amer. Math. Monthly*, pages 227–233, 1938.

[23] D. Lichtblau. Half-GCD and fast rational recovery. In *Proc. ISSAC '05*, pages 231–236. 2005.

[24] J. Massey. Shift-register synthesis and bch decoding. *IEEE Transactions on Information Theory*, 15:122–127, 1969.

[25] R. Moenck. Fast computation of GCDs. In *Proc. of the 5th ACM Annual Symposium on Theory of Computing*, pages 142–171. New York, 1973. ACM Press.

[26] N. Möller. On Schönhage's algorithm and subquadratic integer gcd computation. *Math. Comp.*, 77(261):589–607, 2008.

[27] F. Morain. Implementing the Thull-Yap algorithm for computing euclidean remainder sequences. In *Proc. ISSAC '22*, pages 197–205. 2022.

[28] R. Prony. Essai expérimental et analytique sur les lois de la dilatabilité des fluides élastiques et sur celles de la force expansive de la vapeur de l'eau et de la vapeur de l'alkool, à différentes températures. *J. de l'École Polytechnique Floréal et Plairial, an III*, 1:24–76, 1795. Cahier 22.

[29] A. Schönhage. Schnelle Berechnung von Kettenbruchentwicklungen. *Acta Informatica*, 1(2):139–144, 1971.

[30] A. Schönhage. Schnelle Multiplikation von Polynomen über Körpern der Charakteristik 2. *Acta Informatica*, 7:395–398, 1977.

[31] V. Shoup. NTL: a library for doing number theory. 1996. www.shoup.net/ntl.

[32] D. Stehlé and P. Zimmermann. A binary recursive gcd algorithm. In D. Buell, editor, *Algorithmic Number Theory*, pages 411–425. Springer Berlin Heidelberg, 2004.

[33] S. Stevin. *L'arithmétique*. Imprimerie de Christophle Plantin, 1585.

[34] V. Strassen. Gaussian elimination is not optimal. *Numer. Math.*, 13:352–356, 1969.

[35] V. Strassen. The computational complexity of continued fractions. In *Proc. of the Fourth ACM Symp. on Symbolic and Algebraic Computation*, pages 51–67. 1981.

[36] K. Thull and C. K. Yap. A unified approach to HGCD algorithms for polynomials and integers. https://cs.nyu.edu/yap/papers/SYNOP.htm#hgcd.

[37] A. L. Toom. The complexity of a scheme of functional elements realizing the multiplication of integers. *Soviet Mathematics*, 4(2):714–716, 1963.